\documentclass[aps,
twocolumn,
showpacs,amsmath,amssymb,
superscriptaddress,longbibliography, 10pt]{revtex4-1}
\usepackage{hyperref}

\usepackage{amsmath}
\usepackage{amsfonts}
\usepackage{amssymb}
\usepackage{graphicx}
\graphicspath{{./figures/}}
\usepackage{epsfig}
\usepackage{color}
\usepackage{empheq}
\usepackage{braket}
\usepackage{array}
\usepackage{epstopdf}
\newcommand{\fref}[1]{Fig.~\ref{#1}}

\definecolor{gold}{RGB}{215,155,0}
\definecolor{blue}{RGB}{0,0,255}
\definecolor{darkgreen}{RGB}{20,150,10}
\definecolor{darkblue}{RGB}{10,10,150}
\definecolor{orange}{RGB}{200,100,0}

\newcommand{\MS}[1]{{#1}}
\newcommand{\LM}[1]{{#1}}
\newcommand{\AN}[1]{{#1}}
\newcommand{\COK}[1]{{#1}}
\newcommand{\DM}[1]{{#1}}

\begin{document}
\title{Nonreciprocal transport \MS{based on} cavity Floquet modes in optomechanics}

\author{Laure Mercier de L\'epinay}
\email[]{laure.mercierdelepinay@aalto.fi}
\affiliation{QTF Centre of Excellence, Department of Applied Physics, Aalto University, FI-00076 Aalto, Finland}
\author{Caspar F. Ockeloen-Korppi}
\affiliation{QTF Centre of Excellence, Department of Applied Physics, Aalto University, FI-00076 Aalto, Finland}
\author{Daniel Malz}
\affiliation{Max-Planck-Institut f\"ur Quantenoptik, Hans-Kopfermann-Strasse 1, D-85748 Garching, Germany}
%
%
%
%
%
%
\author{Mika A.~Sillanp\"a\"a}
\affiliation{QTF Centre of Excellence, Department of Applied Physics, Aalto University, FI-00076 Aalto, Finland}

\begin{abstract}
Directional transport is obtained in various multimode systems by driving multiple, non-reciprocally-interfering interactions between individual bosonic modes.
However, systems sustaining the required number of modes become physically complex.
In our microwave-optomechanical experiment, we show how to configure nonreciprocal
transport between frequency components of a single superconducting cavity coupled to two drumhead oscillators. The frequency components are promoted to Floquet modes and generate the missing dimension to realize an isolator and a directional amplifier.
A second cavity left free by this arrangement is used to cool the mechanical oscillators and \LM{bring the transduction noise close to the quantum limit}. 
We furthermore uncover a new type of instability specific to nonreciprocal coupling. Our approach is generic and 
\LM{can greatly simplify }
quantum signal processing and \LM{the design of }topological lattices from low-dimensional systems.
\end{abstract} 
\maketitle 




\textit{Introduction.-- } 
While lattices with a high level of complexity exist in nature, \LM{building} \MS{these in a bottom-up fashion} while aiming on specific functionalities remains challenging. Doing so, however, would be highly desirable in order to utilize topological phenomena in applications. \MS{The construction of artificial lattices} has therefore been investigated in cold atoms \cite{Goldman2016}, photonics \cite{Lu2014}, superconducting circuits \cite{Koch2010,Tsomokos2010,Kollar2019}, and cavity optomechanics \cite{Peano2015,Schmidt2015}. 
An intriguing possibility that has received attention recently to demonstrate complex functionalities \LM{in a physically low-dimensional system is to complement it with synthetic dimensions \cite{Ozawa2019}}. \MS{Initially}, atoms' internal degrees of freedom were identified as lattice sites \cite{Celi2014} aligned along a non-spatial, synthetic dimension supplementing existing spatial dimensions. Floquet quasienergy levels that emerge in a periodically driven nonlinear system \cite{Rechtsman2013,Martin2017,Lin2018,Baum2018,Crowley2019,Peterson2019,Dutt2019}, as well as multiple or degenerate resonant modes \cite{Andrijauskas2018,Lin2018,Chen2019} have also been considered as lattice sites in additional dimensions. Interestingly, nontrivial topology can be designed in these synthetic dimensions just as in spatial dimensions, which, \MS{among many other phenomena, can lead} to nonreciprocal transport \cite{Wanjura2019}.

We focus on microwave cavity optomechanics  \cite{Regal2008,Teufel2011} where microwave resonators interact with mechanical vibrations. 
Microwave-optomechanical signal processing either reciprocal \cite{Massel2011,OckeloenKorppi2016,OckeloenKorppi2017,Toth2017} or nonreciprocal \cite{Malz2018,Peterson2017, Bernier2017, Barzanjeh2017, MercierdeLepinay2019} \COK{shows some advantages} 
over Josephson-junctions-based processing \cite{Movshovich1990,CastellanosBeltran2008,Abdo2013,Zhong2013,Abdo2014,Eichler2014,Sliwa2015,Lecocq2017,Westig2018}\COK{:} 
saturation powers are orders of magnitude higher and 
superconductivity is not \MS{fundamentally} necessary. Multimode optomechanical nonreciprocal devices have recently been suggested \cite{Hafezi2012,Metelmann2015,XuXW2015,  XuXW2016,Li2017,Jiang2018,Malz2018,JiangY2018} and demonstrated  \cite{Hua2016,Shen2016,Ruesink2016, Fang2017, Shen2018, Ruesink2018}. Progress in this direction has nonetheless been hindered by the difficulty of fabricating devices with multiple mechanical modes coupled equivalently and \DM{strongly} enough to multiple electromagnetic modes. 
The \LM{use} of a single cavity mode \LM{for several simultaneous operations }
has been considered for passive detection of stronger processes \cite{Lecocq2015b,Wollman2015,OckeloenKorppi2018}. \AN{Kerr-type nonlinearities have also been shown to promote coupling between Floquet modes \cite{Qiu2019}.} However, exciting a single cavity mode so that driven Floquet components actively participate in the dynamics has received little attention, with the notable exception of Ref.~\cite{Xu2019} where the phase difference of \LM{two} components of a single cavity field is used to realize nonreciprocal mechanical noise transport.

In this work, we \DM{show that} configurable \LM{and} directional electromagnetic-signal \AN{transmission} can be obtained in an optomechanical system 
\AN{by designing a loop of interactions in the synthetic plane generated by}
driven Floquet modes \LM{on one hand} and 
\MS{multiple mechanical modes}
\LM{on the other hand} \COK{to realize a microwave isolator and a directional amplifier}. The use of Floquet modes thus \MS{demonstrated provides a way} to simplify \LM{these nonreciprocal }\AN{devices} and alleviate \MS{practical} requirements. 



\begin{figure*}[ht]
\includegraphics[width=17cm]{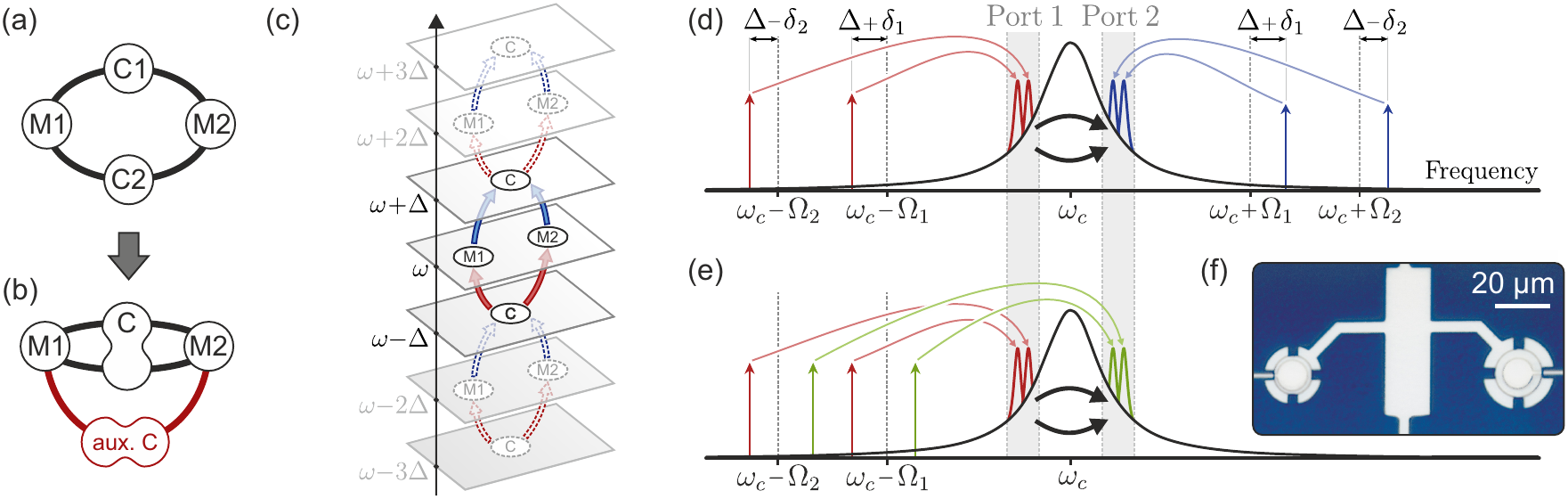}
\caption{\emph{Principle of Floquet nonreciprocal devices.} (a) Physical scheme of a four-mode system \LM{used to implement nonreciprocal \AN{transduction}. (b) Physical scheme of a simplified system exhibiting nonreciprocal transport between off-resonant components of a single cavity field.} An auxiliary cavity mode can be used to sideband-cool mechanical modes \DM{by pumping once again away from exact sidebands not to induce parasitic coupling between mechanical oscillators}. (c) Scheme of the dual mechanically-mediated coupling between Floquet modes lattice sites represented in the generated synthetic dimensions. \AN{Filtering by mechanical susceptibilities restrict the coupling picture to the central plaquette.} (d) Drive frequencies used to build an optomechanical directional amplifier. The auxiliary cavity pumping scheme is not represented \cite{supplement}. (e) Similar driving scheme to build an isolator. (f) Optical micrograph of the pair of micromechanical drum oscillators.}
\label{Fig1}
\end{figure*}

\textit{Principle.-- }  
Let us first consider a multimode cavity optomechanical system as shown on \fref{Fig1}(a) \LM{similar to the ones used in a number of directional \AN{transduction} demonstrations}  \cite{Bernier2017,Barzanjeh2017,Peterson2017,MercierdeLepinay2019}. It has two mechanical oscillators that are both coupled to two cavity modes. Both cavities are \LM{excited with} several coherent pumping tones. Each set of tones enhances one mechanically-mediated coupling mechanism between the cavities, and the relative phase of  pump tones controls \MS{the interference between these two coupling processes}.

\MS{\LM{We demonstrate that }this physical system can be recast into the one on \fref{Fig1}(b) where different frequency components \LM{of one of the cavity fields belonging to a driven Floquet system} play the role of the two cavity modes. The second cavity is left available for auxiliary optomechanical manipulations.} \LM{The mediating mechanical modes still participate at their respective resonance frequencies [see \fref{Fig1}(c)] and their narrow bandwidths play an essential role by restricting the number of Floquet manifolds coupled together.}

\MS{
Let us \LM{first} consider \LM{the configuration of} the Floquet directional amplifier. The pertaining pump angular frequencies} are as schematized in \fref{Fig1}(d): $\left\{\omega_c \pm\left(\Omega_i+\delta_i+\Delta\right)\right\}_{i=1,2}$. We define $\omega_c/2\pi$ the frequency of the cavity,  $\Omega_i/2\pi$ of the mechanical oscillator $i$, the linewidths $\gamma_i$ of the oscillators and the detuning of the pumps $\Delta$ much larger than the mechanical linewidth but smaller than the cavity linewidth $\kappa$. The mechanical damping rates $\gamma_i$ are here assumed to already include auxiliary optical damping. The detunings $\delta_i$ \DM{are} comparable to $\gamma_i$ \DM{and} allow to drive mechanical susceptibilities out of resonance.


\LM{To} realize \LM{instead} a Floquet-mode isolator, 
\COK{all four pump tones are placed close to the red sidebands} 
and the pumps angular frequencies become [see \fref{Fig1}(e)]  $\left\{\omega_c -\left(\Omega_i+\delta_i\pm\Delta\right)\right\}_{i=1,2}$. In both devices, the pumps drive components of the cavity field $\pm\Delta$ away from resonance. The frequency ranges \LM{around these detunings} \LM{play the roles of the two ports of either device instead of cavity modes} [see \fref{Fig1}(d)]. The equations of evolution for electromagnetic and mechanical operators, linearized and with fast rotating terms ignored, display time-dependencies that cannot be eliminated by moving to a rotating frame. For example, for the isolator in the frame rotating with $H_0=\hbar\omega_c a^\dagger a + \sum_i \hbar (\Omega_i+\delta_i) b_i^\dagger b_i$ where $a^\dagger, a$ are photonic creation and annihilation operators, and $b_i^\dagger, b_i$ analogous phononic operators for oscillator $i$, \DM{the Langevin equation read}:
\begin{equation}
\begin{array}{*3{>{\displaystyle}l}}
\dot{a} &\!=& i\sum_j \left(G_{j-}e^{i\Delta t} +G_{j+} e^{-i\Delta  t}\right)b_j -\frac{\kappa}{2}a+ a_{\rm drive}\\
\dot{b}_j &\!=&\! i\delta_j b_j + i\!\sum_j\! \left(G_{j-}^*e^{-i\Delta  t} +G_{j+}^*e^{i\Delta  t}\right)a -\frac{\gamma_j}{2}+b_{j\rm drive}.
\end{array}
\label{eq:eqmotion}
\end{equation}
%
%
$G_{j\pm}= g_{j}\alpha_{j\pm}$ is the enhanced optomechanical coupling for the pump detuned by $\pm \Delta$ associated to mechanical mode $j$, \DM{where $g_{j}$ is the single-photon optomechanical coupling of mode $j$ to the cavity and $\alpha_{j\pm}$ the cavity field \LM{amplitude} at the frequency of the corresponding pump.} The terms $a_{\rm drive}=\sqrt{\kappa_e}a_{\rm in}^e+ \sqrt{\kappa_i}a_{\rm in}^i$ and $b_{j\rm drive}=\sqrt{\gamma_j}b_{j,\rm in}$ model respectively the fluctuating and coherent probe drives of the cavity and the fluctuating drive of mechanical oscillator $j$. 
The total cavity linewidth is the sum of the external and internal loss rates $\kappa=\kappa_e+\kappa_i$. We introduce cooperativities for each pump $\{C_{j\pm}\}_{j=1,2}$ and the only relevant phase degree of freedom $\varphi$ \cite{Malz2018} between pumps, such that $G_{1-} = \sqrt{\gamma_{1} \kappa C_{1-}/4}\,e^{+i\varphi/2}$, $G_{2-} = \sqrt{\gamma_{2} \kappa C_{2-}/4}\,e^{-i\varphi/2}$ and $G_{j+} = \sqrt{\gamma_{j} \kappa C_{j+}/4}$ for $j=1,2$. 
\AN{The phase $\varphi$ is a crucial parameter as it determines the nonreciprocal nature of the coupling and therefore the directionality of the transduction.}
Since $\gamma_{j}$ is by far the smallest frequency scale of the system, \DM{narrow} mechanical susceptibilities  restrict the number of relevant harmonics \DM{to two only} at detunings $-\Delta$ and $+\Delta$  from the cavity resonance frequency \cite{supplement}. These Floquet components define the two ports of the device, thereafter named respectively port 1 and 2.

Eliminating phononic operators from Fourier transformed Eqs.~(\ref{eq:eqmotion}), it follows that a 2-vector of cavity operators is invariant under the evolution equations: $A(\omega)=\begin{pmatrix} a(\omega-\Delta)& a(\omega+\Delta)\end{pmatrix}^T$ in the case of the isolator and $A(\omega)=\begin{pmatrix} a(\omega-\Delta)&a^\dagger(\omega-\Delta)\end{pmatrix}^T$ in the case of the amplifier. 
Defining a global cavity susceptibility \DM{$2\times2$ matrix} $\chi(\omega)$ \cite{supplement}, the vector $A$ is related to similarly-defined drive vectors $A_{\rm in}^e$ and $A_{\rm in}^i$ by: $A(\omega)=\chi(\omega)\left[\sqrt{\kappa_e}A_{\rm in}^e(\omega)+ \sqrt{\kappa_i}A_{\rm in}^i(\omega)\right]$. \LM{Using an analogous definition for the cavity output rate $A_{\rm out}$, the input-output relation \DM{reads} $A_{\rm out}=A_{\rm in}^e-\sqrt{\kappa_e} A$. Therefore, the scattering matrix $S$ defined by $A_{\rm out} = S \,A_{\rm in}^e$ (\DM{temporarily} omitting noise terms) is $S(\omega) = \mathbb{I}_2- \kappa_e \chi(\omega)$.}
The elements $S_{ij}$ of this matrix  define the \DM{scattering} parameters between ports $j$ and $i$. 
%
%

%
\begin{figure*}[ht]
  \centering \includegraphics[width=17.5cm]{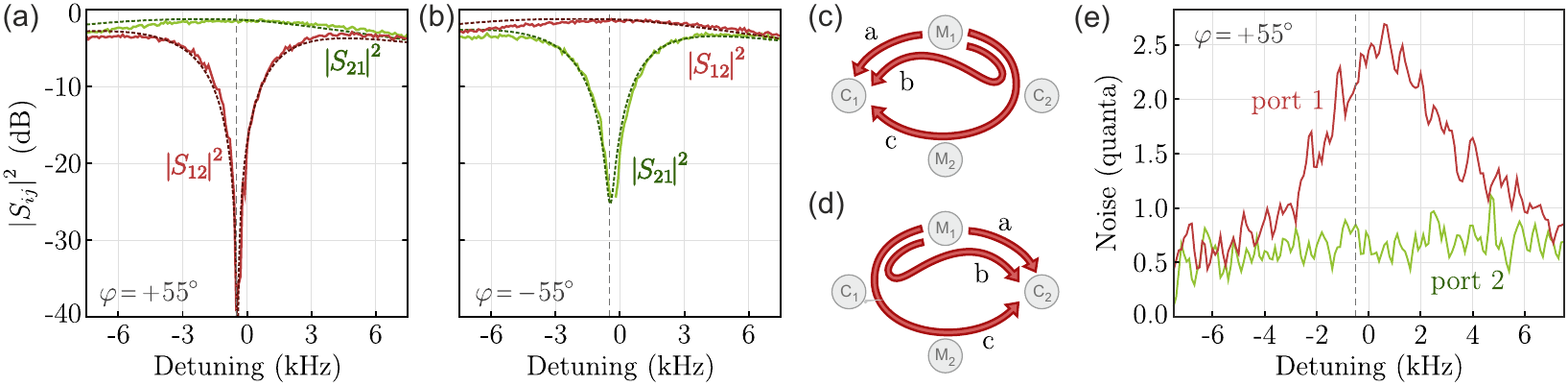}
  \caption{\emph{Microwave isolator.} (a) \DM{Scattering} parameters $S_{12}$ (solid red line) and $S_{21}$ (light green solid line) for $\varphi=55^\circ$ and fits (darker dashed lines). The dashed grey line marks the optimal working frequency. Inset: reflection $S_{11}$ on isolated port. (b) Same as (a) for $\varphi=-55^\circ$. Inset: reflection $S_{22}$ on isolated port. (c) Schematic of the paths taken by noise from mechanical oscillator 1 to port 1. (d) Same as (c) for noise from mechanical oscillator 1 to port 2. (e) For $\varphi=+55^\circ$, noise at port 1 (solid red) corresponding to backward-propagating noise for the isolator with this phase and at port 2 (solid light green) corresponding to forward-propagating noise.}
  \label{Fig2}
\end{figure*}
%
%

\textit{Experiment.-- } An on-chip microwave $LC$ circuit lithographied in aluminum sustains two electromagnetic cavity modes. We coin these ``primary'' and ``auxiliary'' modes. The former is used to establish nonreciprocal transfer and the latter to sideband-cool the mechanical modes \cite{Schliesser2008} in order to reduce noise and broaden the bandwidth. The cavity modes have respective frequencies $\omega_c/2\pi= 4.98\,\rm GHz$ and $\omega_c^a/2\pi= 6.62\,\rm GHz$, and internal and external decay rates $\kappa_{e}/2\pi=1.3\,\rm MHz$, $\kappa_{i}/2\pi=190 \,\rm kHz$ and $\kappa_{e}^a/2\pi=900\,\rm kHz$, $\kappa_{i}^a/2\pi=580 \,\rm kHz$. The circuit includes two vacuum-gap capacitors [see \fref{Fig1}(f)] whose top plate is allowed to move freely, materializing two mechanical oscillators of frequencies $\Omega_1/2\pi = 6.69\,\rm MHz$ and $\Omega_2/2\pi = 9.03\,\rm MHz$ and intrinsic decay rates $\gamma_1^0/2\pi=55\,\rm Hz$ and $\gamma_2^0/2\pi=110\,\rm Hz$. The chip is mounted in a dilution cryostat to be operated at a temperature of  $10\,\rm mK$.



\textit{Isolator.-- }  
\LM{We} now discuss the configuration shown in \fref{Fig1}(e) that employs only red-sideband tones. The \LM{global} susceptibility matrix in the basis defined by $A$ is modified by a coupling matrix $T$:
\begin{equation}
\chi^{-1}(\omega) = \begin{pmatrix} \frac{\kappa}{2} - i(\omega-\Delta) & 0\\
0 & \frac{\kappa}{2} - i(\omega+\Delta) 
\end{pmatrix} + \begin{pmatrix} T_{11} & T_{12}\\
T_{21} & T_{22}
\end{pmatrix}. 
\end{equation}
Diagonal coupling terms $T_{11}$ ($T_{22}$) account for  standard backaction of mechanical modes on Floquet cavity modes from the two pumps \LM{detuned by} $-\Delta$ ($+\Delta)$. Off-diagonal terms $T_{ij}$ ($i\neq j$) also involve one contribution from each mechanically-mediated coupling path between Floquet cavity modes:
\begin{equation}
\begin{array}{*5{>{\displaystyle}l}}
T_{12}(\omega) &=&  G_{1-}G_{1+} \chi_{m,1}(\omega)  &+& G_{2-}G_{2+} \chi_{m,2}(\omega)   \\[6pt]
T_{21}(\omega)  &=& G_{1-}^*G_{1+}^* \chi_{m,1}(\omega)   &+& G_{2-}^*G_{2+}^*\chi_{m,2}(\omega)
\end{array}
\label{eq:couplings}
\end{equation}
where $\chi_{m,j}$ is the susceptibility of mechanical mode $j$, centered on $-\delta_j$: $\chi_{m,j}(\omega)=\left[\gamma_{j}/2-i(\omega+\delta_j)\right]^{-1}$. Off-diagonal elements of the scattering matrix $S_{12}$ and $S_{21}$ are proportional to $T_{12}$ and $T_{21}$ respectively \cite{supplement}. Therefore, to obtain e.g.~isolation of port 1 ($S_{12}=0$), it suffices to cancel out $T_{12}$. In order to maintain simultaneous transfer in the other direction, one must ensure that $S_{21}$, and therefore $T_{21}$, is \LM{concurrently} as high as possible. \AN{This asymmetry is made possible thanks to the phase-shift of each coupling path provided by the off-resonance participation of either mechanical oscillator \cite{supplement}.}



%

The experiment is prepared by sideband-cooling mechanical oscillators through the auxiliary cavity down to $n_{\rm 1 eff} \simeq 4.0$ and $n_{\rm 2 eff} \simeq 8.9$ quanta. Corresponding effective mechanical linewidths are $\gamma_{1}/2\pi \simeq 1.7\,\rm kHz$ and $\gamma_{2}/2\pi \simeq 1.9\,\rm kHz$. \LM{The primary cavity is pumped with} detuning $\Delta/2\pi = 30\,\rm kHz$, much larger than the effective mechanical damping rates. With cooperativities $\{C_{1-},\,C_{1+},\,C_{2-},\,C_{2+}\} = \{ 3.9,\,2.5,\,3.7,\,2.4\}$ and additional detunings $\delta_1/2\pi = -\delta_2/2\pi=1\,\rm kHz$, we show on \fref{Fig2}(a) an optimal isolation for $\varphi=+55^\circ$ of $|S_{12}|^2\simeq-39.3\,\rm dB$ with $|S_{21}|^2\simeq-1.3\,\rm dB$ insertion loss. We also demonstrate for the opposite phase $\varphi=-55^\circ$ a device working in the reversed direction on \fref{Fig2}(b) with $|S_{21}|^2\simeq-24.5\,\rm dB$ isolation and $|S_{12}|^2\simeq-1.2\,\rm dB$ insertion loss. The bandwidth around $3\,\rm kHz$ is comparable to the effective mechanical linewidths. \LM{Due to relatively small mechanical frequency separation, pumps also excite the mechanical mode they are not intended to drive, leading to dynamical backaction taken into account in the theoretical fits presented throughout the paper.}


The  noise in the device arises mainly from mechanical thermal noise \cite{Malz2018} which propagates through 3 paths as indicated on \fref{Fig2}(c): path ``a'' is the direct conversion of phonons into the same amount of cavity photons. The two others (``b'' and ``c'') follow the same route as signals across the device and interfere destructively at the isolated port. Therefore, only path ``a'' contributes to the backward-propagating noise which is thus 
simply half of the mechanical oscillators' total occupation: $n_{\rm back}=\frac{1}{2}(n_1+n_2+1)$ in the limit of high cooperativities and ideal cavity. Here we maintain this noise around $n_{\rm back}  \simeq 2.5$ quanta [see \fref{Fig2}(e)]. By comparison, without sideband-cooling,
the expected input-port noise is $35$ photons in \DM{the} ideal case. At the other port [see \fref{Fig2}(d)],  direct path ``a'' interferes somewhat destructively with the sum of the indirect paths ``b'' and ``c'' and mitigates total fluctuations, which results in \COK{very} small output noise of thermal origin, and thus $n_{\rm out}\simeq0.6$ quanta [see \fref{Fig2}(e)]. 

\begin{figure}
  \centering \includegraphics[width=8.6cm]{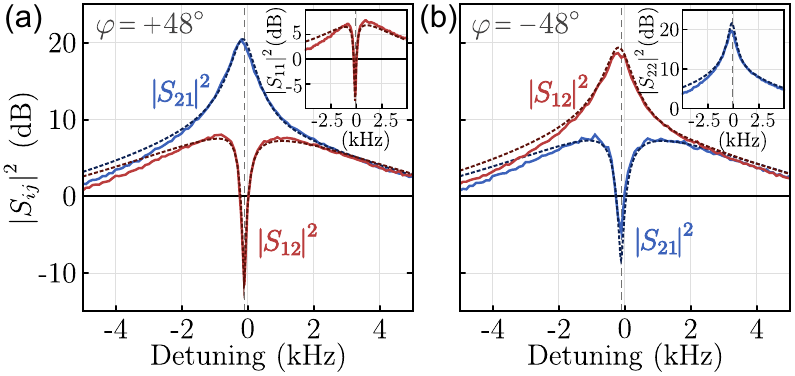}
  \caption{\emph{Directional amplification.} (a) \DM{Scattering} parameters $S_{12}$ (solid red line) and $S_{21}$ (solid blue line) for $\varphi=+48^\circ$ and fits with expressions of the text (darker dashed lines). Inset: reflection $S_{11}$ (solid line) on isolated port and fit (darker dashed line). (b) Same as (a) for the opposite phase $\varphi=-48^\circ$. Inset: reflection $S_{22}$ on isolated port (solid line) and fit (darker dashed line).}
  \label{Fig3}
\end{figure}
%

\textit{Directional amplifier.-- } Owing to blue-sideband driving, the S-parameters of the amplifier relate $a(\omega-\Delta)$ to $a^\dagger(\omega-\Delta)$: they exchange quadratures between input and output ports. This translates the phase-preserving but phase-conjugating nature of the device \cite{Caves82} and entails that signals sent at a frequency $\omega_c-\Delta-\nu$ are converted at $\omega_c+\Delta+\nu$ \cite{MercierdeLepinay2019}, that is, the output frequency is mirrored around the port's central frequency.
As long as the detunings $\Delta$ are small compared to $\kappa$, the gain $\mathrm{G}$ in the limit of ideal cavities and large cooperativities remains the same as for separate-cavity amplifiers \cite{Malz2018, MercierdeLepinay2019, supplement}:
\begin{equation}
\mathrm{G} = \frac{4 C_{1-}C_{1+}}{(C_{1-}-C_{1+})^2}.
\end{equation}

With \LM{similar} pre-cooling of mechanical modes \LM{($n_1\simeq2.9,\,n_2\simeq 8.1,\,\gamma_1/2\pi \simeq 1.6{\,\rm kHz}, \,\gamma_2/2\pi\simeq 0.9{\,\rm kHz}$)} and \LM{same} detuning $\Delta$ as for the isolator, we use cooperativities $\{C_{1-},\,C_{1+},\,C_{2-},\,C_{2+}\} \simeq \{ 4.2,\,3.2,\,5.3,\,2.6\}$ to demonstrate in \fref{Fig3}(a) for $\varphi=+48^\circ$ a maximum amplification gain of $\mathrm{G}\simeq |S_{21}|^2 \simeq 20.3\,\rm dB$ and a simultaneous isolation of $|S_{12}|^2 \simeq -11.7\,\rm dB$. The amplification and the isolation bandwidths $1.5\,\rm kHz$ and  $1\,\rm kHz$ \COK{respectively} are again comparable to the mechanical linewidths, but lower than those of the isolator since they are not enhanced by parasitic coupling  \cite{supplement}. \fref{Fig3}(b) also shows a gain of $|S_{12}|^2 \simeq 18.5\,\rm dB$ and a simultaneous isolation of $|S_{21}|^2 \simeq -4.7\,\rm dB$ with the opposite phase $\varphi=-48^\circ$. However, contrary to the case of the isolator, only one port can be impedance-matched to the transmission line \DM{due to the asymmetric pumping of red sidebands of one Floquet mode and blue sidebands of the other} \cite{supplement}. As a result of this asymmetry, regardless of the phase, port 1 displays low reflectivity $|S_{11}|^2$ and port 2 a large one $|S_{22}|^2$ [see insets to \fref{Fig3}(a) and (b)]. 
The optimal configuration is therefore $\varphi=+48^\circ$ which suppresses power reflected on the input port by $|S_{11}|^2 \simeq -7.4\,\rm dB$.

\begin{figure}
  \centering \includegraphics[width=8.6cm]{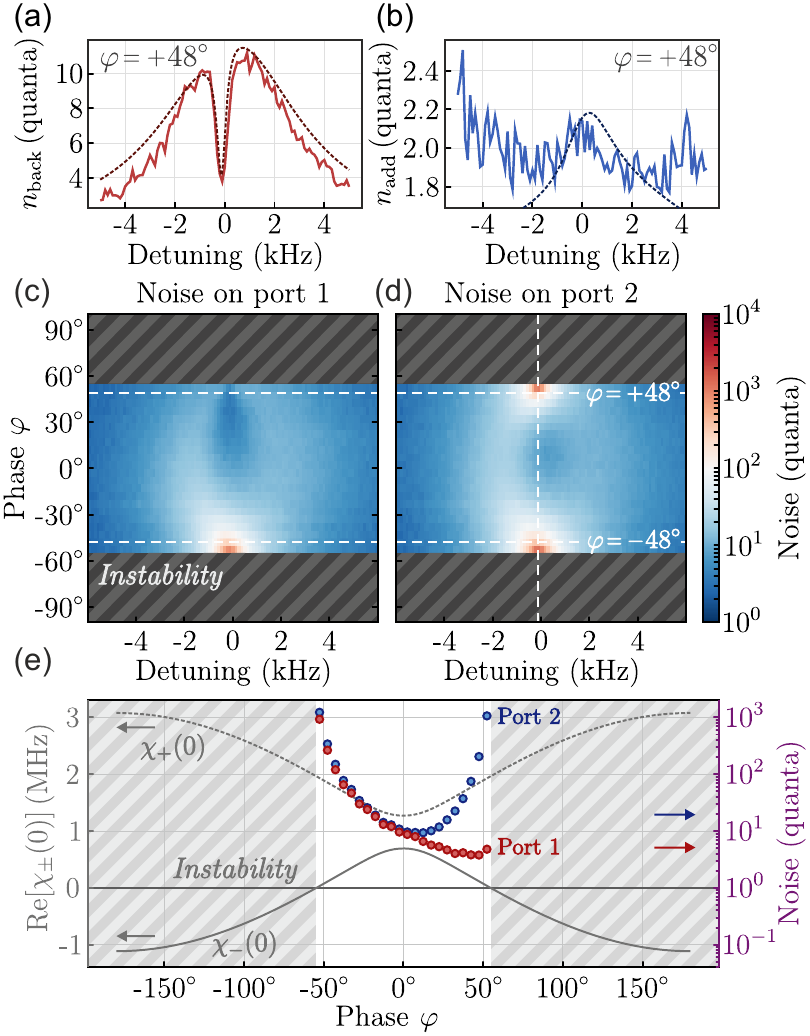}
  \caption{\emph{Noise and phase-dependent instability.} 
   (a) Backward-propagating noise for optimal phase  $\varphi=+48^\circ$, corresponding to noise at port 1 (solid red) and theory superimposed (darker dashed line). (b) Added noise for optimal phase  $\varphi=+48^\circ$ (solid blue) and theory (darker dashed line).
   (c) Noise at port 1 as a function of frequency and pump phase. Grey dashed area: unstable region. Optimal phases $\varphi=\pm 48^\circ$ are shown as white horizontal dashed lines. (d) Same as (c) but at port 2. Dashed vertical white line: frequency of optimal directionality. (e) Calculated real-parts of the optical eigen-susceptibilities at zero frequency (solid and dashed grey lines, right-side scale). The lower real-part is negative for $|\varphi|>55^\circ$, causing the instability. Noise at optimal frequency [cut of (d) along dashed vertical line] at port 2  (blue dots, left-side scale) diverges at the onset of the instability, as does noise at port 1 (red dots, left-side scale) although noise paths' destructive interference at $\varphi=+48^\circ$ limits the maximum measured noise.}
  \label{Fig4}
\end{figure}

In contrast to the isolator, fluctuations propagating across the amplifier are amplified, which results in a generally high noise. However, we measure in this configuration \DM{only} $n_{\rm back} \simeq 3.8$ quanta at port 1 when it is isolated \DM{at} $\varphi=+48^\circ$  [see \fref{Fig4}(c)]. 
Indeed, amplified paths ``b'' and ``c'' \LM{again} interfere destructively and only the non-amplified noise from direct path ``a'' contributes to the backwards noise which is again, in the high cooperativities limit, half of the total mechanical occupancy and is reduced thanks to active cooling. \COK{Output noise results from the same three-path interference as in the isolator and at phase $\varphi=+48^\circ$ the added noise is $n_{\rm add}=n_{\rm out}/\mathrm{G}  \simeq 2.1$ \LM{photons} [see \fref{Fig4}(b)], close to the quantum limit of 0.5 photons.}
On the other hand, for $\varphi=-48^\circ$, port 2 which displays high reflective gain $S_{22}$ \COK{outputs}\LM{ a backward-propagating noise reaching} $n_{\rm back}  \simeq 360$ quanta even with aggressive auxiliary cooling. This asymmetry between noise at phases $\varphi=\pm 48^\circ$ is visible on \fref{Fig4}(c) and (d).
This represents a second reason, together with asymmetric impedance-matching, for which the behavior of the device is not inverted by simply changing the sign of the phase. 


\textit{Phase-dependent instability.-- }  Nonreciprocal amplification furthermore reveals a type of instability for a range of phases. Contrary to optomechanical instability, it does not arise directly from a strong blue-sideband driving but is related to the emergence of an unstable eigenmode of the coupling matrix $\chi$.
Figures \ref{Fig4}(c) and (d) show the amplifier noise for a limited range of phase parameters $|\varphi|<55^\circ$ because the device is unstable for the rest of the $360^\circ$ range. As shown in \fref{Fig4}(e), the onset of the instability coincides with the phase at which one of the eigenvalues of the electromagnetic susceptibility matrix acquires a negative real-part, which is tantamount to a negative damping rate. Figure \ref{Fig4}(e) furthermore shows that the noise diverges at these phases. In the isolator case, on the other hand, the eigenvalue with lower real-part is stabilized by dynamical backaction and never crosses zero. The observed instability is therefore specific to nonreciprocally coupled, non-stabilized multimode devices. As such, it relates to the instability observed in other phase-preserving nonreciprocal coupling cases  \cite{Gloppe2014}.
%
%

\textit{Conclusion.-- } 
We have theoretically and experimentally demonstrated a new archetype of nonreciprocal optomechanical devices based on the interference of Floquet modes in a single cavity. This physical simplification allows to accommodate auxiliary optomechanical manipulations of mechanical oscillators to closely approach the quantum limit of the  transduction. 
We foresee that this approach can greatly simplify signal processing  in other physical platforms involving resonators.
Finally, we uncovered a class of instability arising in nonreciprocally-coupled systems provided they are not stabilized by dynamical backaction.

\begin{acknowledgments}
\AN{We thank Andreas Nunnenkamp, Clara Wanjura and Matteo Brunelli for useful discussions.} This work was supported by the Academy of Finland (contracts 308290, 307757), by the European Research Council (615755-CAVITYQPD), and by the Aalto Centre for Quantum Engineering. The work was performed as  part of the Academy of Finland Centre of Excellence program (project 312057). We acknowledge funding from the European Union's Horizon 2020 research and innovation program under grant agreement No.~732894 (FETPRO HOT). D.M. acknowledges funding from ERC Advanced Grant QENOCOBA under the EU Horizon 2020 program (Grant Agreement No. 742102). We acknowledge the facilities and technical support of Otaniemi research infrastructure for Micro and Nanotechnologies (OtaNano) that is part of the European Microkelvin Platform.
\end{acknowledgments}


%

\end{document}